\documentclass[journal,twoside,web]{ieeecolor}
\usepackage[T1]{fontenc}

\usepackage{jsen}
\usepackage{cite}
\usepackage{amsmath,amssymb,amsfonts}
\usepackage{algorithmic}
\usepackage{graphicx}
\usepackage{textcomp}
\usepackage{wrapfig}
\usepackage{fontawesome}
\usepackage{pifont}
\usepackage{hyperref}
\usepackage{textcomp}
\def\BibTeX{{\rm B\kern-.05em{\sc i\kern-.025em b}\kern-.08em
    T\kern-.1667em\lower.7ex\hbox{E}\kern-.125emX}}

\makeatletter
\def\ps@headings{%
  \def\@oddhead{}%
  \def\@evenhead{}%
  \def\@oddfoot{}%
  \def\@evenfoot{}%
}
\def\ps@IEEEtitlepagestyle{%
  \def\@oddhead{}%
  \def\@evenhead{}%
}
\makeatother

\definecolor{abstractbg}{rgb}{0.89804,0.94510,0.83137}
\begin{document}

\thispagestyle{empty}
\pagestyle{empty}

\title{Design and Development of Low-Cost Datalogger for Indoor and Outdoor Air Quality Monitoring}
\author{%
\centering
\mbox{Prasannaa Kumar D., Gulshan Kumar, Jay Dhariwal, and Seshan Srirangarajan}
}

\IEEEtitleabstractindextext{%
\fcolorbox{abstractbg}{abstractbg}{%
\begin{minipage}{\textwidth}%

\begin{wrapfigure}[20]{r}{3in}
\includegraphics[width=2.5in]{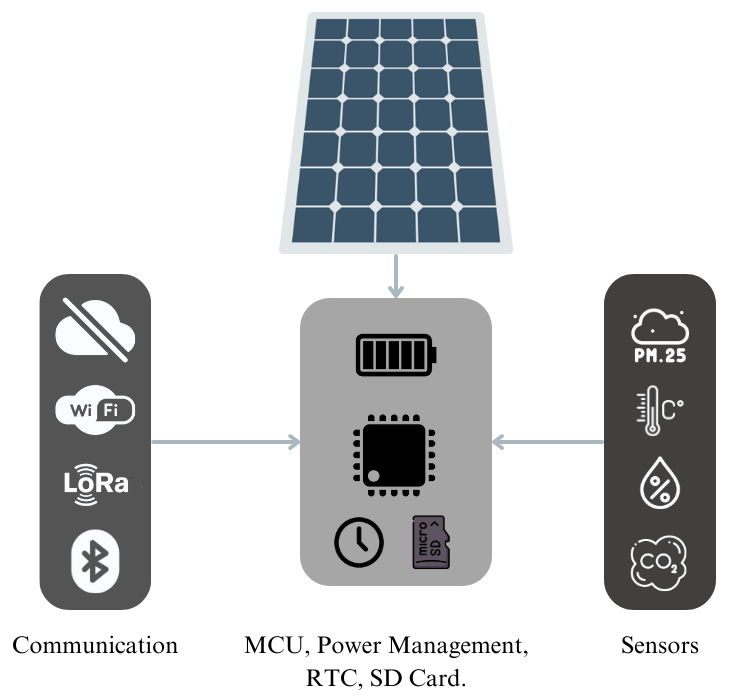}
\end{wrapfigure}

\begin{abstract}
The rising demand for low-cost air quality monitors stems from increased public awareness and interest within the research community. These monitors play a pivotal role in empowering citizens and scientists to comprehend spatiotemporal variations in air quality parameters, aiding in the formulation of effective mitigation policies. The primary challenge lies in the diverse array of application scenarios these monitors encounter. The developed data logging device is exceptionally well-suited for air quality monitoring applications, offering exceptional versatility by seamlessly operating on a range of power sources, including solar energy, batteries, and direct electrical supply. The integration of a built-in battery charger enhances its applicability for deployment in regions with solar power or intermittent electricity availability. To ensure strong network connectivity, the advanced datalogger seamlessly integrates with WiFi, Bluetooth, and LoRaWAN networks. A notable feature is its adaptable MCU system, enabling users to swap the MCU based on specific connectivity, power, and computational requirements. Importantly, the system carefully identifies key parameters crucial for both indoor and outdoor air quality assessment, customizing sensor selection accordingly.

Furthermore, optimization efforts have prioritized energy efficiency, enabling the system to function with minimal power consumption while maintaining data integrity. Additional I2C and UART ports facilitate the monitoring of supplementary parameters.
\end{abstract}


\end{minipage}}}

\maketitle
\thispagestyle{empty}
\section{Introduction}
\label{sec:introduction}
\IEEEPARstart{I}n the world of air quality monitoring, there has been significant progress in the development of Low-Cost monitors, addressing the critical need for improved spatiotemporal resolution compared to conventional monitoring systems ~\cite{Brauer2019,Snyder2013,Morawska2018,Shindler2021}. These monitors find extensive applications in community and personal monitoring, adaptable for both indoor and outdoor environments~\cite{Steinle2015,Bulot2019} due to their compact size and energy efficiency. While they may not match the data precision of reference sensors ~\cite{Forehead2020,Badura2018,Shindler2021}, these budget-friendly alternatives have demonstrated their effectiveness, particularly in non-regulatory contexts~\cite{Duvall2021}.
	
In the context of cost-effective air quality monitoring systems, these devices typically consist of fundamental components including a data logger, sensor array, power supply, and microcontroller unit (MCU). These systems commonly employ a range of sensors to assess air quality, encompassing $\text{PM}_{2.5}$ sensors, temperature sensors, humidity sensors, and carbon dioxide ($\text{CO}_{2}$) sensors.
$\text{PM}_{2.5}$ sensors, often based on the principle of light scattering~\cite{Giordano2021}, provide critical information regarding the concentration of solid airborne particles. The latest USEPA guidelines recommend a $\text{PM}_{2.5}$ (aerodynamic diameter $\leq 2.5~\mu \text{m}$) concentration of $5~\mu \text{g/m}^{3}$. The levels of $\text{PM}_{2.5}$ are greatly influenced by anthropogenic emissions, meteorological conditions and exhibit spatiotemporal variations~\cite{Luo2017,Lin2014,Liu2020}. Extensive research has demonstrated that both prolonged and short-term exposure~\cite{Li2019,Alexeeff2021,Pun2017} to $\text{PM}_{2.5}$ significantly impacts human health. Consequently, continuous monitoring of $\text{PM}_{2.5}$ levels is essential, especially in areas with high $\text{PM}_{2.5}$ concentrations.

$\text{CO}_{2}$ sensors usually utilize electrochemical principles or non-dispersive infrared (NDIR) technology. Maintaining a safe indoor $\text{CO}_{2}$ level is paramount, typically ranging from 400 to 1000 ppm. However, indoor $\text{CO}_{2}$ concentrations can escalate swiftly due to factors such as high occupancy, inadequate ventilation, elevated outdoor levels, and combustion processes like heating or cooking. Numerous studies have highlighted that $\text{CO}_{2}$ levels inside naturally ventilated structures can easily surge to a range of 2500 to 5000 ppm~\cite{DiGilio2021,Franco2020,Becerra2020}. Consequently, ongoing monitoring is vital to uphold optimal indoor air quality in such environments. 

Temperature and humidity sensors detect alterations in resistance or capacitance to measure their respective parameters. These sensors hold significant importance in monitoring indoor and outdoor air quality as they enhance the accuracy of $\text{PM}_{2.5}$ sensors~\cite{Kumar2021,Hofman2020} and aid in calculations related to thermal comfort~\cite{Kumar2014,Valinejadshoubi2021}.

The MCU selection for these devices depends on parameters like power supply, communication, and computational needs. Power sources and internet connectivity are typically easier to get indoors however, for remote applications, battery powered or solar powered systems are preferred.

The primary challenge associated with these monitoring devices lies in their adaptability across various application scenarios. For instance, deploying air quality monitors in remote areas necessitates a solar-powered system with offline data logging~\cite{Kelleher2018}. Conversely, urban deployments are best served with direct power and WiFi connectivity~\cite{Connolly2022}. In certain applications, LoRaWAN is the recommended choice~\cite{Johnston2019}, while battery-powered systems are essential for specific conditions~\cite{Dam2017}. Furthermore, continuous connectivity with an offline data logger may be required for some purposes. The wide variety of application scenarios necessitates a resilient system with extensive customization capabilities to cater to each unique use case.

In this study, we developed a novel data-logging solution designed to accommodate all potential use cases. The developed system is versatile, capable of operating on battery power, solar energy, or direct power supply. It incorporates an integrated charging system for batteries and is compatible with various power sources. The data logger is equipped with three I2C ports and one UART port, offering flexibility in sensor selection and utilization. Additionally, the system is optimized to minimize power consumption. Notably, the system stands out due to its capacity for MCU interchange, enabling adaptation based on the specific requirements of connectivity and computational power.

\section{Design of Datalogger}

\begin{figure}[!t]
\centerline{\includegraphics[width=\columnwidth]{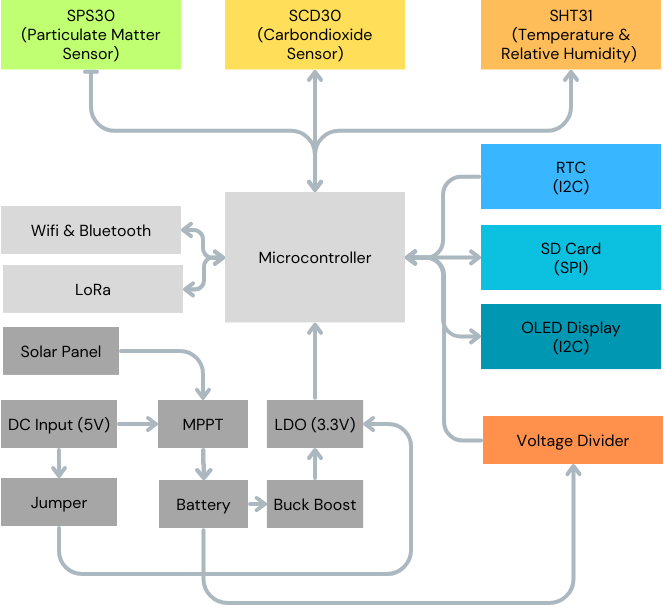}}
\caption{Block diagram of the modular datalogger for air quality monitoring.}
\label{fig1}
\end{figure}

\begin{figure}[!t]
\centerline{\includegraphics[width=\columnwidth]{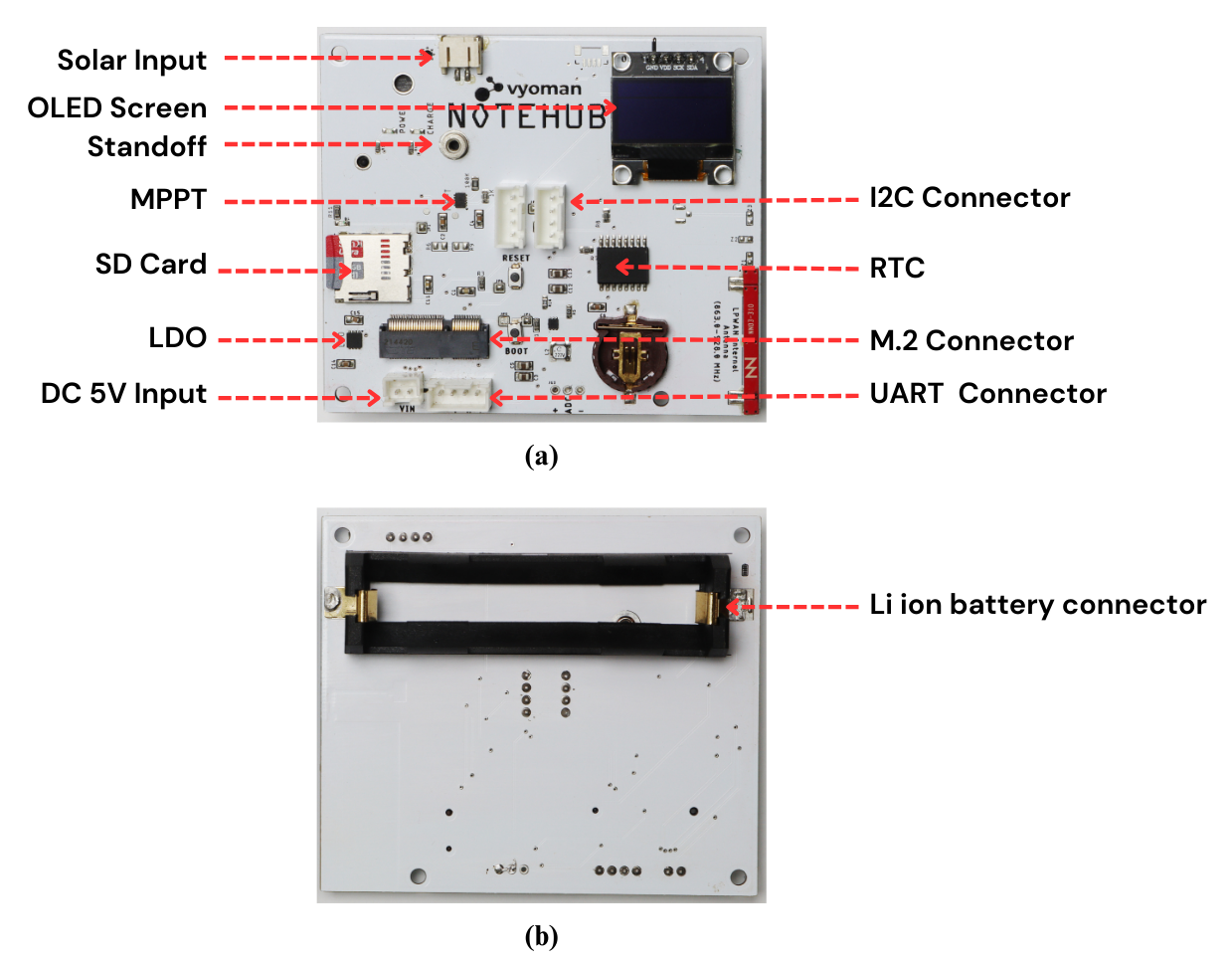}}
\caption{Datalogger baseboard: (a) top view with components (b) bottom view with 18650 battery holder.}
\label{fig2}
\end{figure}

This section outlines the design and development of the datalogger board, encompassing both its hardware and software components. The interchangeable MCU board plays a central role in managing various elements, including sensors, real-time clock (RTC), SD card, and OLED display. Fig.~\ref{fig1} provides a broad illustration of the datalogger's hardware components, while Fig.~\ref{fig2} presents the fabricated datalogger base board, showing both the top side with all components soldered and the bottom side with the $18650$ battery connector.

\subsection{Selection of Microcontrollers}
The selection of the MCU was based on the use of open-source hardware and firmware that allows the achievement of the low-cost objective of the end device. The MCU were selected based on features like low power consumption, and wireless radio availability. The datalogger consists of an M.2 connecter to host the MCU PCB (refer Fig.~\ref{fig3}(a)). Three MCU boards were developed for the datalogger for features like offline data logging, WiFi-based data communication and LoRaWAN-based wireless communication.

\subsubsection{ESP32}
The ESP32 MCU (refer Fig.~\ref{fig3}(b)) is known for its energy efficiency, which is crucial for dataloggers that need to operate on battery power for extended periods. It has various power-saving modes and can be programmed to wake up at specific intervals to record data and then go back to sleep, minimizing power consumption. ESP32 comes with built-in Wi-Fi and Bluetooth capabilities. This allows dataloggers to transmit data wirelessly to a central server or a user's device, making it convenient for remote monitoring and data retrieval. ESP32 has a dual-core processor and a reasonable amount of RAM, which makes it capable of handling data processing tasks, such as data filtering, compression, and encryption, before storing or transmitting data. 

ESP32 can be interfaced with various storage options, including SD cards, SPI flash memory, or even cloud-based storage solutions, making it versatile for different datalogging applications. ESP32 offers a wide range of peripheral interfaces, including GPIO pins, I2C, SPI, UART, and analog inputs. This allows dataloggers to interface with a variety of sensors and data acquisition devices. ESP32 is programmable using popular languages like Arduino, MicroPython, and C/C++. This flexibility makes it easier for developers to create custom datalogging applications tailored to their specific needs. \cite{alfano2020}

The ESP32 has a large and active community of developers and a wealth of open-source libraries and resources available. This makes it easier to find support, share knowledge, and leverage existing code for datalogging projects. ESP32 modules are relatively affordable, making them a cost-effective choice for datalogger development, especially for hobbyists and small-scale projects. ESP32 modules are compact and can be integrated into small enclosures or devices, making them suitable for portable and space-constrained datalogging applications. ESP32 includes hardware-based security features like secure boot and flash encryption, which can be important for protecting the integrity of data collected by dataloggers, especially in sensitive applications.

In summary, the ESP32's combination of low power consumption, wireless connectivity options, processing power, and extensive peripheral support makes it a popular choice for dataloggers in a wide range of applications, from environmental monitoring to industrial data collection.

\subsubsection{STM32G070KBT6}
The STM32G070KBT6 (refer Fig.~\ref{fig3}(c)) is one of the best suitable MCU for an offline datalogger. The STM32G070KBT6 is designed for low-power operation, crucial for an offline datalogger that might run on battery power for extended periods without external power sources. It provides sufficient Flash memory and RAM for storing and buffering data locally, which is essential for offline data logging where data might be collected over time and then periodically saved. The MCU supports various communication interfaces, including UART, SPI and I2C, which can be used to interface with sensors or peripherals for data acquisition. It features multiple GPIO pins that can be configured as analog or digital inputs/outputs, making it versatile for interfacing with a wide range of sensors and data sources.

The MCU offers a low standby current, which helps preserve battery life when the datalogger is in idle mode. It is available in compact packages, making it suitable for space-constrained offline datalogger designs.

\subsubsection{STM32WL Series}

\begin{figure}[!t]
\centerline{\includegraphics[width=\columnwidth]{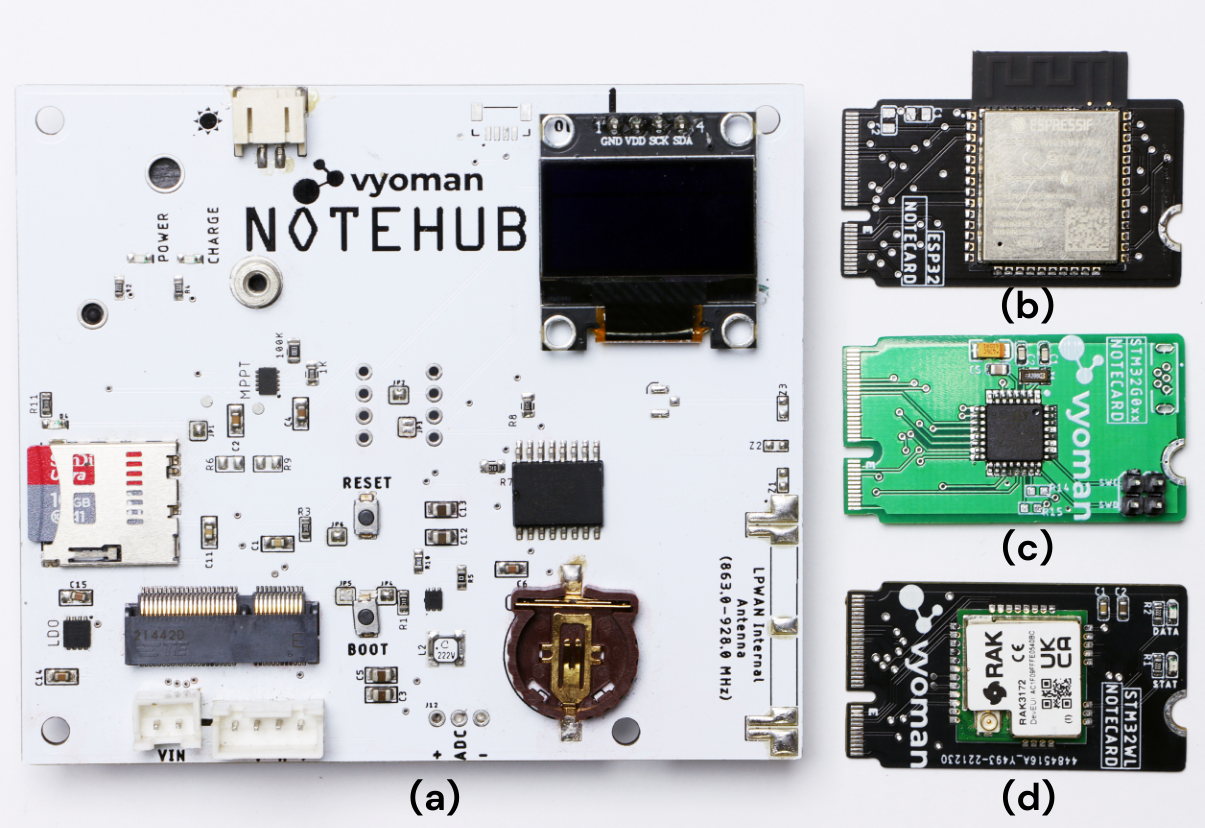}}
\caption{{Microcontroller modules and datalogger baseboard: (a) baseboard (b) ESP32 module (c) STM32G070KBT6 module (d) STM32WLE5CC module.}}
\label{fig3}
\end{figure}

The STM32WL is a series of MCU developed by STMicroelectronics that are specifically designed for applications that require long-range wireless communication. A module from RAK Wireless (RAK3172) based on STM32WLE5CC (refer Fig.~\ref{fig3}(d)) was selected. One of the key features of the STM32WL series is its support for long-range communication using sub-GHz radio frequencies (e.g., 868 MHz and 915 MHz). This makes it well-suited for dataloggers that need to transmit data over extended distances, such as agricultural monitoring in large fields or remote environmental data collection. STM32WL MCU are designed to be energy-efficient, which is important for dataloggers that need to operate on battery power for extended periods. They offer low-power modes and features like low-power RF communication, allowing for longer battery life.

 The STM32WL integrates RF transceivers, including support for LoRa modulation, which is commonly used for long-range IoT communication. This integration simplifies the hardware design and reduces the need for external components. STM32WL MCU are based on the Arm Cortex-M4 core, which provides sufficient processing power for dataloggers that require data processing, data compression, or other computational tasks before storing or transmitting data. STM32WL devices include security features such as hardware encryption, secure boot, and secure firmware updates. This is crucial for ensuring the integrity and confidentiality of data collected by dataloggers. STMicroelectronics provides a comprehensive development ecosystem for STM32 MCU, including software development tools, libraries, and community support. This can simplify the development process for datalogging projects. Additionally, packages for Arduino IDE and Mbed OS are available with an active community of developers.

\subsection{Selection of Sensors}

\subsubsection{Particulate Matter Sensor}
The light scattering principle forms the basis for low-cost $\text{PM}_{2.5}$ sensors. Typically, these sensors feature a structure comprising an airflow pathway, a scattering chamber, a Photodiode (PD) for detecting scattered light, and a MCU to convert this scattered light into electrical signals~\cite{alfano2020}. Calibration of these sensors is conducted using reference sensors to enhance their accuracy. Among the particulate matter pollutants, $\text{PM}_{2.5}$ is major concern due to its ability to deeply penetrate the lungs, resulting in severe health implications~\cite{kim2015}. In this study, the Sensirion SPS30 was selected as the $\text{PM}_{2.5}$ sensor. The SPS30 has been widely adopted in various studies, demonstrating a strong correlation with reference sensors, prolonged sensor functionality, and minimal intermodal variability~\cite{kuula2020,sousan2021}. Notably, the SPS30 is equipped with a sleep mode~\cite{sps30}, rendering it highly suitable for solar-powered and battery-operated devices.

\subsubsection{\texorpdfstring{$\text{CO}_{2}$}{CO2} Sensor}
For $\text{CO}_{2}$ monitoring, the Sensirion SCD30 has been selected as the optimal choice. Extensive evaluations have demonstrated the SCD30's exceptional correlation with the reference sensor across both indoor and outdoor conditions~\cite{tryner2021,mujan2021,macagga2023}. The operational temperature span of 0 to 50 degrees Celsius and humidity ranging from 0 to 95\% allows versatile utilization of this device across various environments. The sensor is capable of measuring $\text{CO}_{2}$ levels up to 40,000 ppm using both I2C and UART communication methods. During sampling, it maintains an average power consumption of just 19mA, peaking at 75mA~\cite{SCD30}. Such efficiency makes it exceptionally well-suited for applications powered by energy-efficient solar sources or batteries.

\subsubsection{Temperature and Humidity sensor}
Although the SCD30 is equipped with an integrated Temperature and Humidity sensor (SHT40), this study utilized the SHT31 due to its superior response time and extended range for both Temperature and Humidity. The SHT31 is connected to the datalogger board via the I2C connection port, as several applications do not necessitate these specific parameters.
\subsection{Power Management}
\subsubsection{Solar/DC Charger}
BQ24210DQCR, a battery charger IC manufactured by Texas Instruments has been used in the datalogger for charging the battery with either solar or DC (5V) power. While it may not be a core component of a datalogger, it plays a crucial role in the power management and functionality of the datalogger, especially if the datalogger operates on battery power. The BQ24210DQCR is used to efficiently charge and manage the state of the battery. It provides features such as constant current and constant voltage charging, which are essential for safely and efficiently charging lithium-ion or lithium-polymer batteries. Lithium-ion batteries can be sensitive to overcharging, over-discharging, and overcurrent conditions. The BQ24210DQCR includes safety mechanisms like thermal shutdown, short-circuit protection, and under-voltage lockout to ensure the battery is charged safely, extending its lifespan and reducing the risk of battery-related accidents. BQ24210DQCR typically includes status pins that can be used to monitor the charging status and control the charging process. When the datalogger is not actively charging the battery, the IC typically has a low standby current, which helps conserve power, important for battery-powered dataloggers.
\subsubsection{DC Step-up boost}
MAX17224 is used to step up and provide constant voltage (5V) from a lithium-ion battery to the entire datalogger. The 18650 lithium-ion battery typically has a nominal voltage of 3.7 volts. However, other datalogger components, such as MCU and sensors, require voltages between 3.3V to 5V for proper operation. The MAX17224 boosts the battery voltage to a 5V, regulated voltage level, ensuring that the datalogger operates correctly as the battery voltage drops over time. The MAX17224 is designed for high efficiency, which is crucial for battery-powered dataloggers. It minimizes power losses during the voltage-boosting process, helping to extend the battery life and make the most efficient use of the available energy in the 18650 battery. When the datalogger is operating in a low-power state, the MAX17224 typically has a low quiescent current.
\subsubsection{Low-dropout voltage regulator}
The NCV8165ML330TCG is a voltage regulator IC designed to provide a stable output voltage from a higher input voltage source. The 5V output from the step-up buck is regulated into 3.3V as most of the components in the datalogger work with a 3.3V input supply. Voltage fluctuations or variations in the input voltage can adversely affect the operation of sensitive electronic components. This voltage regulator helps stabilize the power supply voltage, ensuring consistent and reliable performance of the datalogger.

\subsubsection{Voltage divider}
A simple voltage divider is used to monitor the voltage of the battery.

\subsection{Firmware}
\begin{figure}[!t]
\centerline{\includegraphics[width=\columnwidth]{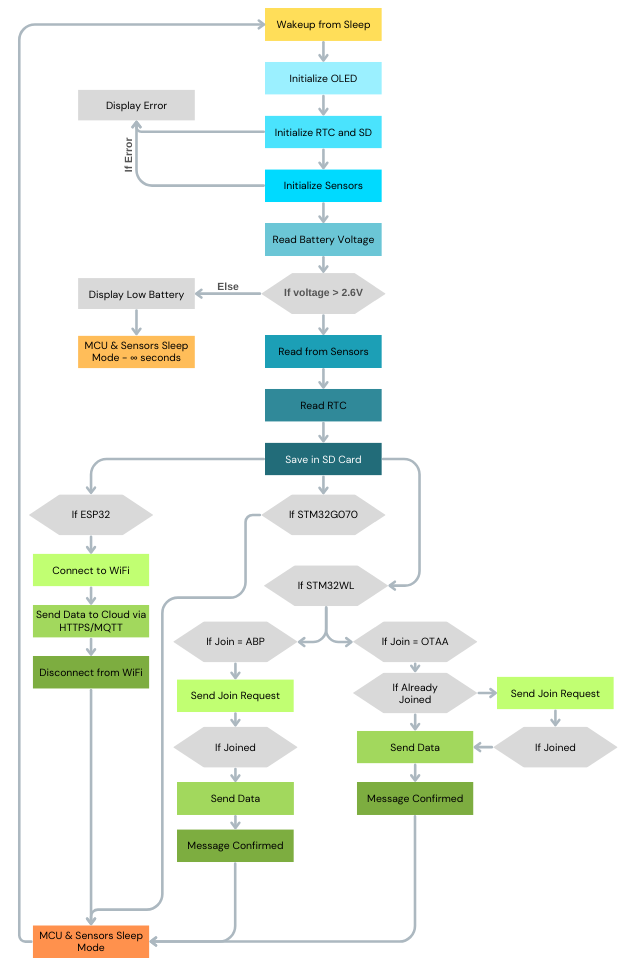}}
\caption{Firmware flowchart of the datalogger operation.}
\label{Firmware_flow}
\end{figure}

The firmware controls the operation, including data acquisition, storage, and retrieval of the datalogger. Fig.~\ref{Firmware_flow} shows the firmware flow common for all the 3 MCU. Upon power-up or wake-up from sleep mode, the battery voltage (if powered by battery) is read using the voltage divider. If the nominal voltage is above 2.6V, the firmware initializes the data logger's hardware components, such as RTC, SD card and sensors. Settings like sampling rates, measurement parameters, storage options, and communication protocols are also initialized. If the battery’s voltage is below 2.6V, the MCU and the sensors are set to sleep mode until charged.  

The firmware continuously or periodically collects data from connected sensors. The interval of data collection depends on the user’s interest and battery ratings. The firmware typically associates timestamps with each data point to record when the measurement was taken. These timestamps are crucial for analyzing and interpreting the data later. The firmware manages the storage of acquired data in the logger's memory. Data can be stored in various formats, such as raw data values, timestamped readings, or aggregated statistics. The data is also displayed in a small OLED display upon interest. Depending upon the MCU, the data can be uploaded to the cloud via WiFi or LoRaWAN. To conserve power and extend the logger's operating life, the firmware uses power-saving features, such as sleep modes or low-power states. The firmware flow in a data logger is highly dependent on the logger's specific design, intended use, and features. Upon requirement, the firmware needs to be modified.

\section{Evaluation and Tests}
\subsection{Experiments}

\begin{figure}[!t]
\centerline{\includegraphics[width=\columnwidth]{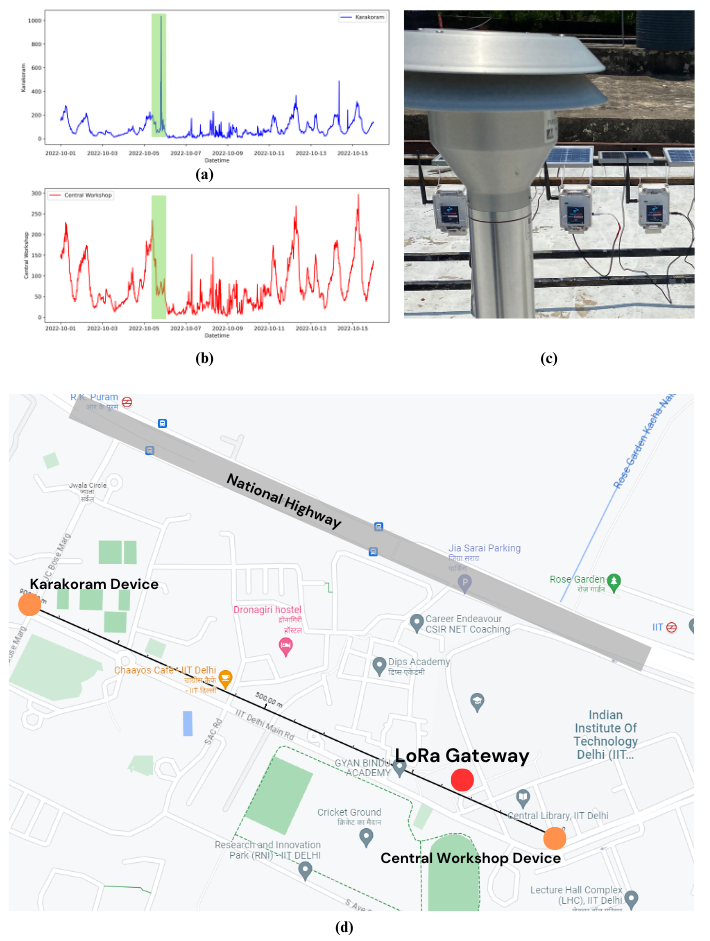}}
\caption{{Sensor network deployment: (a) Karakoram PM$_{2.5}$ data (b) Central Workshop PM$_{2.5}$ data (c) collocated datalogger setup with reference BAM (d) deployment locations and LoRa gateway.}}
\label{sensor_network_locations}
\end{figure}

Two sensor network devices were made based on the datalogger discussed above. The boards were encased in an IP64 enclosure with an SPS30 sensor mounted on it based on the guidelines provided by the manufacturer. The device sends data to the LoRaWAN gateway every 15 minutes once. The device joins the LoRaWAN network using over the air activation (OTAA) method. Device addresses are assigned to every device on new sessions unlike (Activation by Personalization) ABP where the device address and session keys are fixed. This helps in routing the packet brokers to different networks/clusters ~\cite{9838306}. The frame counters linked to ABP devices should remain unchanged throughout their operational lifespan; otherwise, there is a risk of messages from these end devices being discarded. In contrast, OTAA end devices engage in the process of renegotiating frame counters and session keys each time they establish a new session. Consequently, the longevity of an OTAA device is not dependent on the frame counter's capacity. Additionally, the adaptive data rate (ADR) mechanism was also used in the devices to optimize data rates, airtime and energy consumption ~\cite{9908849}. The ADR helps in optimizing the transmit power, resulting in low power consumption.  

This experiment was conducted as part of a sensor network to examine the spatiotemporal variation of $\text{PM}_{2.5}$ levels. Initially, the sensors were collocated with a beta attenuation monitor (BAM), as shown in Fig.~\ref{sensor_network_locations}(c). Subsequently, the devices were deployed at two distinct locations within the campus, as illustrated in Fig.~\ref{sensor_network_locations}(d). The Karakoram device was installed in a hostel area near a major national highway with minimal obstructions, whereas the Central Workshop device was located within the academic section of the campus. The LoRa gateway was installed atop a five-story building to ensure reliable communication coverage.

Figs.~\ref{sensor_network_locations}(a) and \ref{sensor_network_locations}(b) present the $\text{PM}_{2.5}$ time series recorded by the Karakoram and Central Workshop devices, respectively, over a 15-day period. During the Dusshera event (05-10-2023), involving the burning of effigies within the campus, the Karakoram device recorded a peak concentration of 1036~$\mu \text{g/m}^3$, as highlighted by the shaded region in Fig.~\ref{sensor_network_locations}(a). In contrast, the Central Workshop device did not exhibit a significant peak during this period, as shown in Fig.~\ref{sensor_network_locations}(b). Table~\ref{tab:PM_Comparison} summarizes the device performance, indicating uptime exceeding $98\%$ for both devices.            

\begin{table}[!t]
\caption{Device uptime and PM$_{2.5}$ statistics at deployment locations}
\label{tab:PM_Comparison}
\centering
\renewcommand{\arraystretch}{1.1}
\resizebox{\columnwidth}{!}{
\begin{tabular}{|c|l|c|c|c|c|}
\hline
\textbf{S. No.} & \textbf{Location} & \textbf{Uptime (\%)} & \textbf{Min ($\boldsymbol{\mu}$g/m$^3$)} & \textbf{Max ($\boldsymbol{\mu}$g/m$^3$)} & \textbf{Avg ($\boldsymbol{\mu}$g/m$^3$)} \\
\hline
1 & Central Workshop & 98.26 & 2 & 297 & 72.92 \\
\hline
2 & Karakoram & 99.86 & 2 & 1036 & 81.25 \\
\hline
\end{tabular}
}
\end{table}

\subsection{Results}
The power profiler kit 2 (PPK2) from Nordic Semiconductor is a useful tool for measuring the power consumption of electronic devices. The primary purpose of measuring power consumption is to optimize the energy efficiency of electronic devices like dataloggers. By accurately quantifying how much power a datalogger consumes during various operating states, it is easy to identify areas for improvement and reduce unnecessary power usage. This leads to longer battery life.

\begin{figure}[!t]
\centerline{\includegraphics[width=\columnwidth]{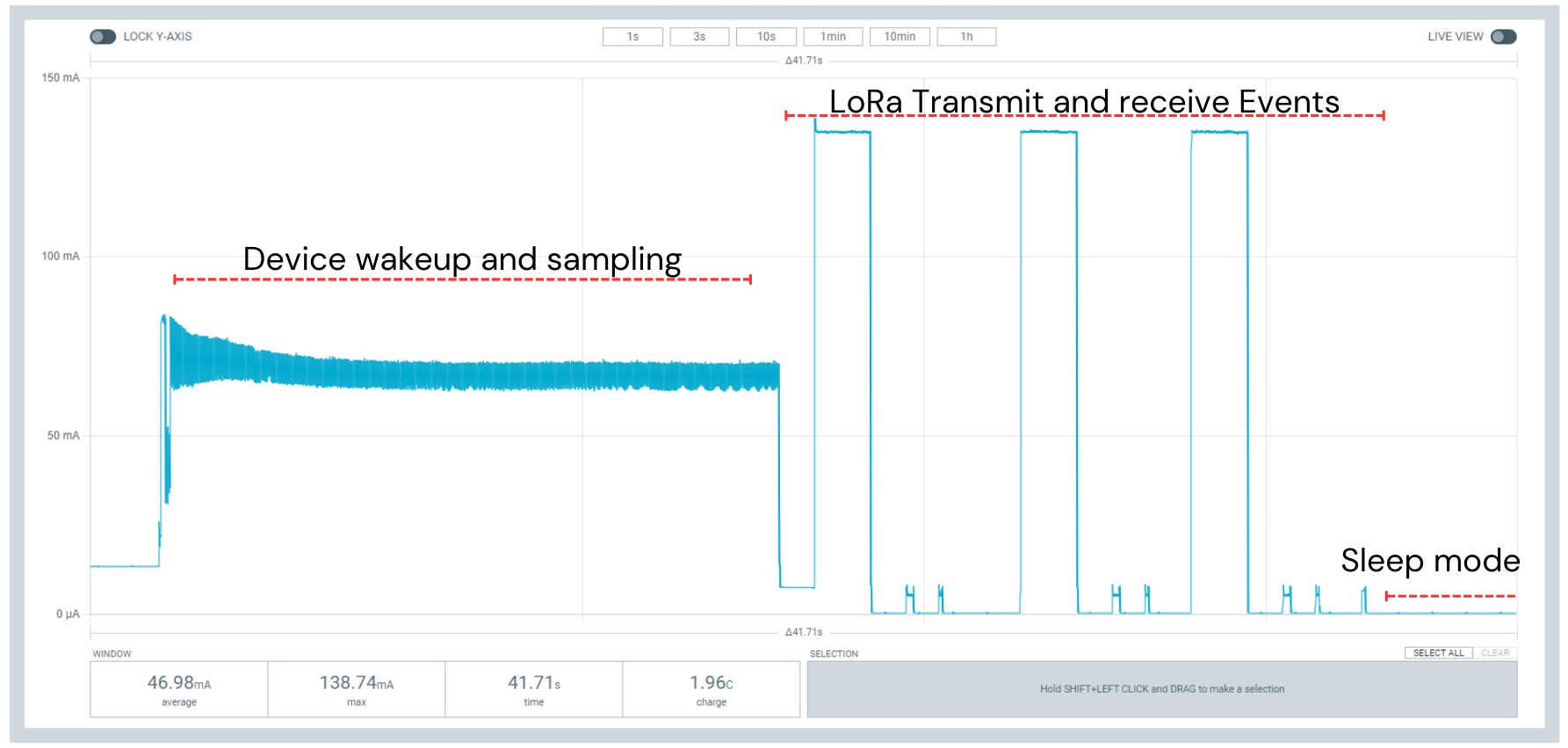}}
\caption{Power consumption profile of the STM32WL-based LoRaWAN device during wake-up, data transmission/reception, and sleep modes for PM$_{2.5}$ monitoring.}
\label{Power_profile_stm_PM}
\end{figure}

\begin{figure}[!t]
\centerline{\includegraphics[width=\columnwidth]{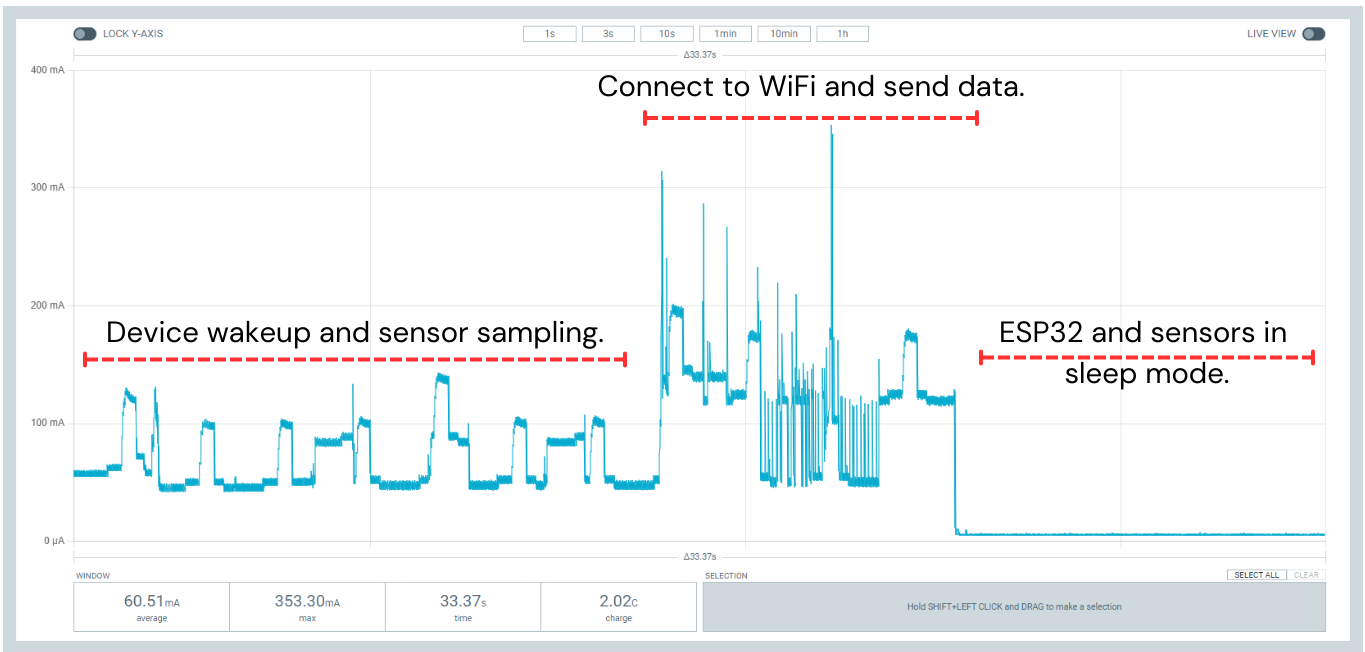}}
\caption{Power consumption profile of the ESP32-based indoor device during sensor sampling, WiFi data transmission, and sleep modes for CO$_2$ monitoring.}
\label{power_co2}
\end{figure}

Fig.~\ref{power_co2} illustrates the power consumption of an indoor device designed for ($\text{CO}_{2}$) monitoring. This device is equipped with the ESP32 MCU and features the SCD30 and SHT31 sensors. The sensors perform three sampling operations, each with a 2-second interval between samples. Following the sampling process, the data is time-stamped and stored on the SD card. After this storage step, the ESP32 connects to a WiFi network and transmits the data using the HTTPS protocol. Upon successful data transmission, the sensors are switched to sleep mode, and the ESP32 enters deep sleep mode for a duration of 4 minutes. The device remains active for a total of 25 seconds, with an average power consumption of 60mA, as depicted in Fig.~\ref{Power_profile_stm_PM}. Once the device enters sleep mode, it consumes only an average current of 7mA.

Fig.~\ref{Power_profile_stm_PM} shows the power consumption of an outdoor device designed for ($\text{PM}_{2.5}$) monitoring. The device uses RAK3172, an STM32WL-based MCU. SPS30 is used for $\text{PM}_{2.5}$ monitoring.    

\subsection{Costs}
The objective of this project was to create a novel datalogger that is affordable, reliable and capable of operating independently. The budget of the datalogger is discussed in Table~\ref{tab:budget_MCU}. The total cost of the datalogger depends on the MCU and sensors used as shown in Table~\ref{tab:budget_sensor}. The budget of a sensor network device used for $\text{PM}_{2.5}$ monitoring costs around $80\$$. When mass-produced, the total costs could be reduced considerably. 

\begin{table}[!t]
\caption{Cost breakdown of the low-cost datalogger hardware}
\label{tab:budget_MCU}
\centering
\footnotesize
\setlength{\tabcolsep}{3pt}
\renewcommand{\arraystretch}{1.0}
\begin{tabular}{|c|p{0.52\columnwidth}|c|}
\hline
\textbf{S. No.} & \textbf{Component} & \textbf{Price (\$)} \\
\hline
1 & Datalogger baseboard & 18 \\
\hline
2 & ESP32 MCU board & 4.5 \\
\hline
3 & STM32WL MCU board & 6 \\
\hline
4 & STM32G070 MCU board & 3.5 \\
\hline
5 & 3300 mAh battery & 2.6 \\
\hline
6 & IP64 enclosure & 5 \\
\hline
\end{tabular}
\end{table}

\begin{table}[!t]
\caption{Cost of sensors used in the datalogger system}
\label{tab:budget_sensor}
\centering
\footnotesize
\setlength{\tabcolsep}{3pt}
\renewcommand{\arraystretch}{1.0}
\begin{tabular}{|c|p{0.52\columnwidth}|c|}
\hline
\textbf{S. No.} & \textbf{Sensor} & \textbf{Price (\$)} \\
\hline
1 & SPS30 & 48.5 \\
\hline
2 & SCD30 & 57.3 \\
\hline
3 & SHT31 & 9 \\
\hline
\end{tabular}
\end{table}

\section{Conclusion}

%https://aneescraftsmanship.com/check-marks%E2%9C%94%E2%9C%98%E2%98%91%E2%98%92%E2%A8%82%E2%93%A5-in-latex/
% \begin{table}[]
% \centering
% \begin{tabular}{|p{15pt}|p{35pt}|p{20pt}|p{20pt}|p{30pt}|p{20pt}|}
% \hline
% S.No & Features & LCDL & Saveris 2-H1 ~\cite{testo} & Waspmote ~\cite{Libelium_2023} & DL-PM ~\cite{Decentlab} \\
% \hline 1 & Modular & \ding{51} & \ding{53} & \ding{51} & \ding{53}  \\
% \hline 2 & WiFi & \ding{51} & \ding{51} & \ding{51} & \ding{53} \\
% \hline 3 & LoRa & \ding{51} & \ding{53} & \ding{51} & \ding{51} \\
% \hline 4 & Offline Logging & \ding{51} & \ding{51} & \ding{51} & \ding{51} \\
% \hline 5 & Solar & \ding{51} & \ding{53} & \ding{51} & \ding{51} \\
% \hline 6 & External power & \ding{51} & \ding{51} & \ding{51} & \ding{53} \\
% \hline 7 & Multiple Sensors & \ding{51} & \ding{53} & \ding{51} & \ding{53} \\
% \hline 8 & Low Cost & \ding{51} & \ding{53} & \ding{53} & \ding{53} \\

% \hline
% \end{tabular}
% \caption{Comparison of Dataloggers available in the market}
% \label{tab:my_label}
% \end{table}

\appendices

A novel datalogger has been developed for the purpose of air quality monitoring. This versatile datalogger is designed to operate using various power sources including direct electrical supply, solar power, and batteries. It offers flexibility in data storage by incorporating an inbuilt SD card and the ability to upload data directly to the web via WiFi, Bluetooth, or LoRaWAN network. The development of this datalogger considered a wide array of scenarios including outdoor and indoor settings, offline conditions, as well as solar and battery-powered operations. Efficient power management has been a focal point during the optimization of the datalogger. Utilizing this technology, two air quality monitoring devices have been deployed within the IIT Delhi campus. These monitors successfully recorded $\text{PM}_{2.5}$ data for a continuous period of 15 days, relying solely on solar and battery power. Notably, the sensor uptime for both monitors exceeded an impressive $98\%$. The implementation of a unique MCU swap system enhances customization capabilities. Furthermore, the datalogger is equipped with additional UART and I2C ports to facilitate the integration of supplementary parameters.
% \section*{Acknowledgment}

\bibliographystyle{IEEEtran}
\bibliography{bibliography}

\end{document}